\documentclass[useAMS,usenatbib]{mn2e}

\usepackage{lscape,graphicx,ulem,amssymb,amsmath}

\def\msol{\rm{M}$_{\odot}$}
\def\lsol{\rm{L}$_{\odot}$}

\def\HII{H\,{\sc ii}}

\def\arcsec{$^{\prime}$$^{\prime}$}
\def\arcmin{$^{\prime}$}
\def\deg{$^{\circ}$}
\def\micron{$\mu$m}
\def\etal{\textit{et al.}}
\def\kms{km\,s$^{-1}$}

\newcounter{ppnum4}
\newcounter{ppnum5}
\title[A cluster of outflows in the Vulpecula Rift]{A cluster of outflows in the Vulpecula Rift}
\author[J.~C.~Mottram and C.~M.~Brunt]{J.~C.~Mottram$^{1}$\thanks{E-mail:
joe@astro.ex.ac.uk} and C.~M.~Brunt$^{1}$\\
$^{1}$School of Physics, University of Exeter, Exeter, Devon, EX4 4QL, UK}
\begin{document}

\date{Accepted 2011 September 15. Received 2011 September 14; in original form 2011 June 16}

\pagerange{\pageref{firstpage}--\pageref{lastpage}} \pubyear{2011}

\maketitle

\label{firstpage}

\begin{abstract}

We present $^{12}$CO, $^{13}$CO and C$^{18}$O (J=3$-$2) observations of a new cluster of outflows in the Vulpecula Rift with HARP-B on the JCMT. The mass associated with the outflows, measured using the $^{12}$CO HARP-B observations and assuming a distance to the region of 2.3~kpc, is 129~\msol{}, while the mass associated with the dense gas from C$^{18}$O observations is 458~\msol{} and the associated sub-millimeter core has a mass of 327~$\pm$~112~\msol{} independently determined from Bolocam 1.1mm data. The outflow-to-core mass ratio is therefore $\sim$0.4, making this region one of the most efficient observed thus far with more than an order of magnitude more mass in the outflow than would be expected based on previous results. The kinetic energy associated with the flows, 94$\times$10$^{45}$~ergs, is enough to drive the turbulence in the local clump, and potentially unbind the local region altogether. The detection of SiO (J=8$-$7) emission toward the outflows indicates that the flow is still active, and not simply a fossil flow. We also model the SEDs of the four YSOs associated with the molecular material, finding them all to be of mid to early B spectral type. The energetic nature of the outflows and significant reservoir of cold dust detected in the sub-mm suggest that these intermediate mass YSOs will continue to accrete and become massive, rather than reach the main sequence at their current mass.

\end{abstract}

\begin{keywords}
ISM: jets and outflows, stars: winds, outflows, molecular data, stars:formation, stars: pre-main-sequence
\end{keywords}

\section{Introduction}
\label{S:intro}

Molecular outflows are observed towards stars of all masses during their formation \citep[][]{Arce2007}, and are generally associated with both active accretion \citep[see e.g.][]{Churchwell1999,Pudritz2007} and loss of angular momentum from the star/disk system that they originate from \citep[e.g.][]{Tomisaka2000,Yamada2009}. Early observations suggested that different scenarios for outflow generation might be required for the low and high mass regimes \citep[see e.g.][and references therein.]{Richer2000}. However, higher-resolution studies of massive outflows have shown that these differences were exaggerated by the low-resolution single-dish observations used and that their properties transition relatively smoothly as a function of mass \citep[e.g.][]{Beuther2002b,Beuther2002c}. 

\begin{figure*}
\center
\includegraphics[width=0.999\textwidth]{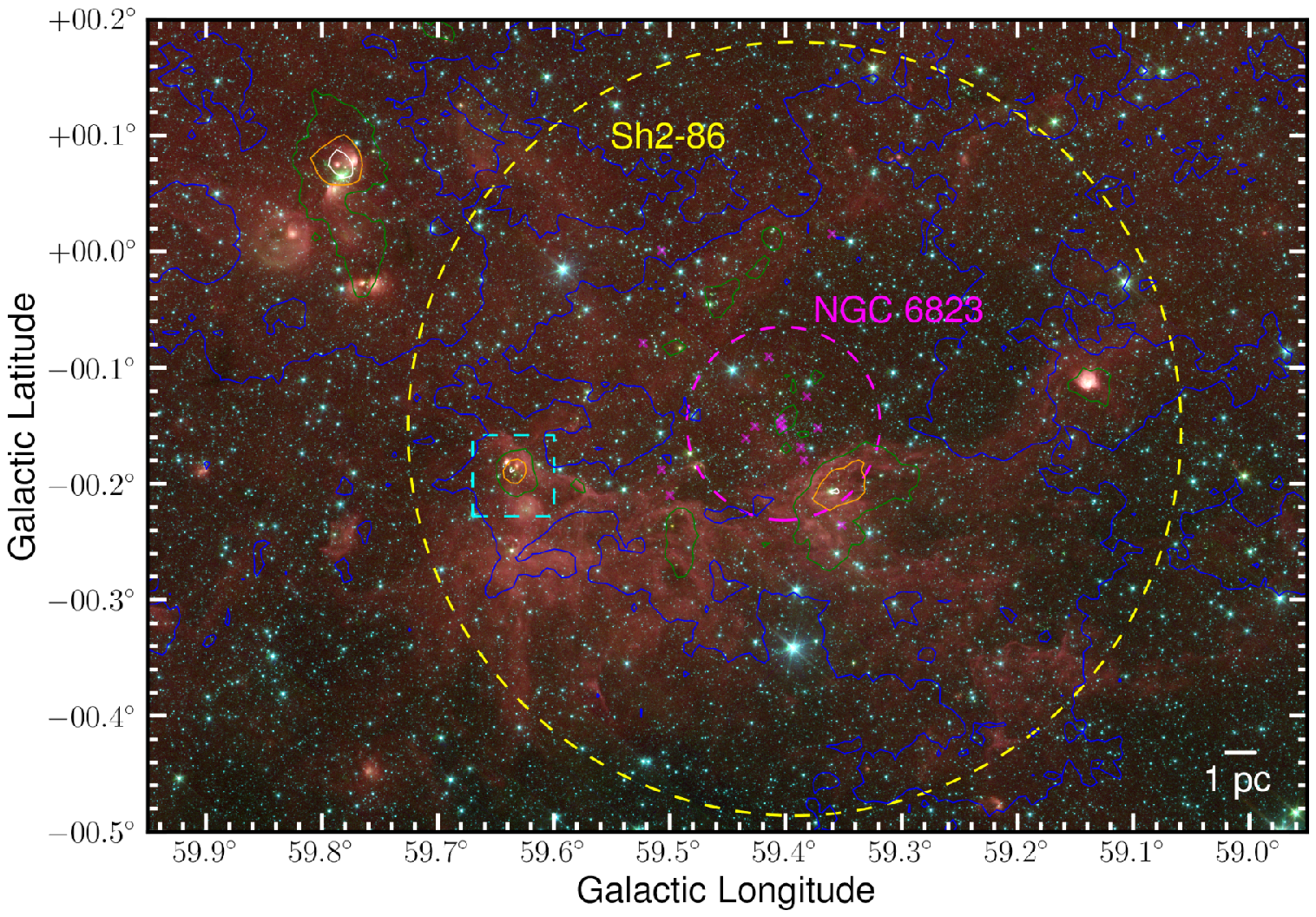}
\caption{Infrared three-colour GLIMPSE \citep[][]{Benjamin2003} IRAC 3.6~\micron{} (blue), 4.5~\micron{} (green) and 8.0~\micron{} (red) image of part of the Vulpecula Rift with overlaid contours showing integrated Exeter-FCRAO Survey \citep[][Brunt et al., 2011, in prep.,]{Mottram2010b} $^{12}$CO emission. The contour levels for the integrated emission are 25, 50, 75 and 100 K\kms{}, and are coloured blue, green, orange and white respectively. The cyan dashed square indicates the region around the newly discovered outflow which will be explored in this paper, while the yellow dashed circle indicates the original identification of the evolved \HII{} region Sh2-86 by \citet{Sharpless1959}. The magenta crosses indicate the positions of the OB stars identified towards the cluster NGC6832 by \citet{Massey1995}, while the magenta dashed circle shows the approximate region containing the main cluster as identified by \citet{Bica2008}.}
\label{F:intro_int12co}
\end{figure*}

In addition to being signposts of active star formation, the mass, energy and angular momentum transported by molecular outflows can also have a significant impact on the surrounding molecular clouds in which young stars form. This can be through the driving of turbulence within the cloud \citep[e.g.][]{Matzner2007,Brunt2009}, at least on small scales, or, for particularly powerful outflows, even contribute to the dispersal of the cloud and removal of the reservoir for further star formation \citep[][]{Matzner2000}. A recent study of the outflows in the Perseus molecular cloud by \citet[][see also \citealt{Hatchell2009} and \citealt{Arce2010}]{Curtis2010} has shown that outflows can contribute a significant amount of energy to driving turbulence and cloud dispersal for whole molecular clouds, though they may not always be energetic enough to dominate these processes on their own. Recent investigations into how the energy and momentum transported by outflows is translated to the cloud \citep[e.g.][]{Banerjee2007,Nakamura2007,Carroll2009} show efficiencies which can be strongly scale-dependent, so even if outflows in a given cloud are not powerful enough to disrupt the whole cloud complex, they may be significant on the intermediate cluster scales.

In this paper we present recent observations of a new outflow discovered at $\ell$~=~59.6375\deg{}, $b$~=~$-$00.1875\deg{} in the $^{12}$CO (J=1$-$0) observations of the Exeter FCRAO (Five-College Radio Astronomical Observatory) CO Survey \citep[][Brunt et al., 2011, in prep.]{Mottram2010b}, which have a spatial resolution of 45\arcsec{} and a velocity resolution of $\sim$0.15~\kms{}. The systematic velocity of the cloud emission associated with the outflow is $\sim$27.5~\kms{}, and the outflow is essentially unresolved in these data. A three-colour GLIMPSE \citep[][]{Benjamin2003} map, with contours showing the integrated FCRAO $^{12}$CO (J=1$-$0) emission in the region and the outflow position marked is presented in Figure~\ref{F:intro_int12co}. 

A sub-mm/mm core identified by \citet[][]{Chapin2008} using BLAST \citep[][]{Pascale2008} at $\ell$~=~059.6331, $b~=~-$00.1906 (their V52), who derive a core mass of 390~$\pm$~75~\msol{} and luminosity of 3850~$\pm$~610~\lsol{}, is associated with the outflow. The core is also detected in the Bolocam Galactic Plane Survey \citep[BGPS][]{Aguirre2011} at 1.1mm which, using the median gas temperature of 27~K (obtained during outflow property calculation in section~\ref{S:properties}), the source integrated flux and equation~10 from \citet{Rosolowsky2010}, gives a core mass of 327~$\pm$~112~\msol{}.

\begin{figure*}
\center
\includegraphics[width=0.98\textwidth]{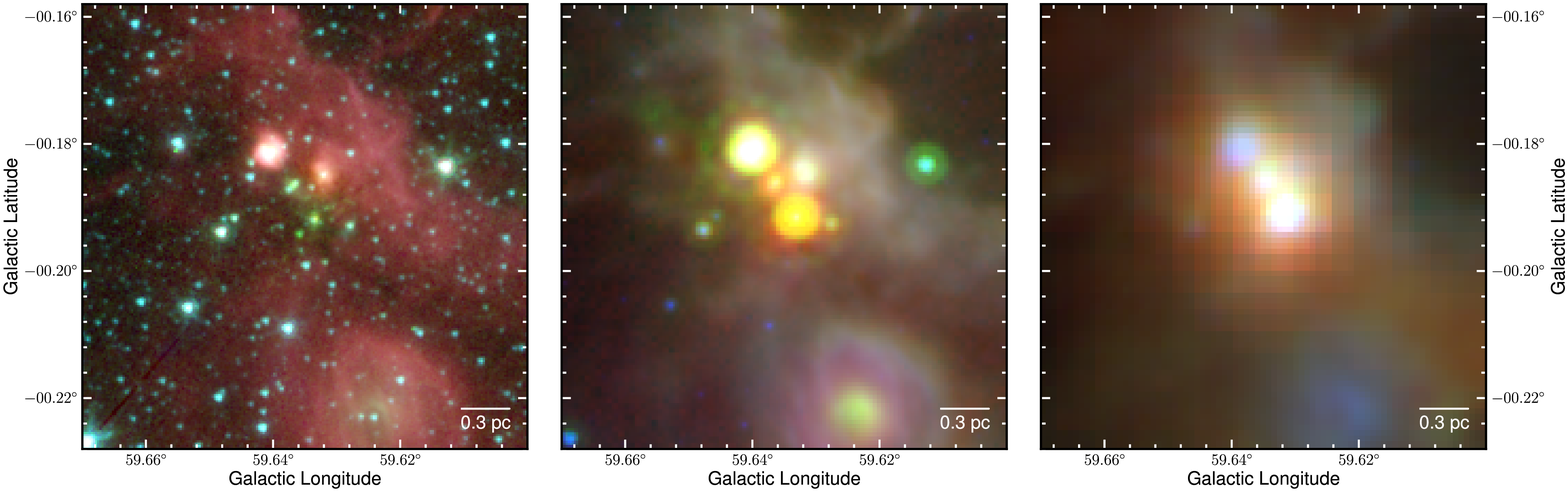}
\caption{Infrared three-colour images of the region around the outflows. Left: GLIMPSE IRAC 3.6~\micron{} (blue), 4.5~\micron{} (green) and 8.0~\micron{} (red). Centre: GLIMPSE IRAC 8.0~\micron{} (blue), MIPSGAL \citep[][]{Carey2009} 24~\micron{} (green) and Hi-GAL \citep[][]{Molinari2010a,Molinari2010b} PACS 70~\micron{} (red). Right: Hi-GAL PACS 70~\micron{} (blue), PACS 160~\micron{} (green) and SPIRE 250~\micron{} (red).}
\label{F:intro_3colour}
\end{figure*}

The outflow also lies near several mid/far-IR bright sources as shown in the three-colour combination GLIMPSE, MIPSGAL \citep[][]{Carey2009} and Hi-GAL \citep[][]{Molinari2010a,Molinari2010b} images in Figure~\ref{F:intro_3colour}, which are part of the Vulpecula Rift molecular cloud complex. The region is near the large \HII{} region Sh-86 \citep{Sharpless1959} and the young cluster NGC 6832 \citep[][see Figure~\ref{F:intro_int12co}]{Massey1995}, part of the VulOB 1 OB association. There have been some suggestions that the whole region is undergoing sequential star formation \citep[e.g.][]{Ehlerova2001}, but \citet{Billot2010} find no evidence for this in the YSO population as obtained from GLIMPSE and MIPSGAL observations. In terms of the early phases of star formation and evolution, \citet{Billot2011} found an increase in clustering of sources as wavelength decreases in the five Herschel bands (70~\micron{}, 160~\micron{}, 250~\micron{}, 350~\micron{}, 500~\micron{}) of the Hi-GAL Science Demonstration Phase (SDP) observations of this region \citep{Molinari2010b}. While detailed identification of evolutionary phases has not been completed, this tentatively indicates that warmer sources are more likely to be in clusters within the region. The molecular cloud properties of the whole Vulpecula Rift as observed in the Exeter FCRAO CO Survey will be discussed in an upcoming paper (Mottram \etal{}, in prep.).

The kinematic distance relating to the systematic outflow velocity, using the \citet{Brand1993} rotation curve, is 2.7~kpc if the source is at the near distance and 5.9~kpc if at the far distance. The outflow is most likely associated with the Vulpecula Rift molecular cloud complex, which infrared extinction maps suggest is at the near distance \citep[e.g.][]{Russeil2011}. The open cluster NGC 6823 also lies within the cloud complex, for which \citet{Massey1995} derive a photometric distance of 2.3~kpc. We therefore follow the approach of both \citet{Chapin2008} and \citet{Russeil2011} in assuming a distance of 2.3~kpc to the region. 

It is perhaps surprising that an outflow associated with such a massive core has not been discovered sooner, but this might be because it only has an upper limit at 25\micron{} in the IRAS point source catalogue \citep[PSC,][]{Beichman1988} which would have caused it to be excluded from the selection criteria of previous studies (e.g. \citealp{Beuther2002b}, who used the sample of \citealp{Sridharan2002}).

We begin by describing the JCMT observations in section~\ref{S:observations}, after which we describe the method used to identify the outflow velocity windows and calculate outflow properties in section~\ref{S:properties}. In section~\ref{S:results} we present results and analysis of the outflow properties, as well as those relating to the local molecular material and the young stellar objects (YSOs) most probably associated with the outflow. Finally, we provide a short discussion of the wider context of our results and reach our conclusions in sect.~\ref{S:conc}.

\section{JCMT Observations}
\label{S:observations}

\begin{table*}
\centering
\caption{Observational Setup}
\begin{tabular}{@{~}ccccccc@{~}}
\hline
\centering
Transition & Rest frequency & Bandwidth & No. of channels & Vel. Resolution & Obs. Type & On source time \\
& (GHz) & (MHz) & & (\kms{}) &  & (min) \\
\hline
CO (J=1$-$0) & 345.80 & 250 & 4096 & 0.053 & Raster & 52 \\
& 345.80 & 1000 & 1024 & 0.847 & Raster & 52 \\
$^{13}$CO (J=1$-$0) & 330.59 & 250 & 4096 & 0.055 & Raster & 71 \\
C$^{18}$O (j=1$-$0) & 329.33 & 250& 4096 & 0.056 & Raster & 71 \\
HCO$^{+}$ (J=4$-$3) & 356.73 & 250 & 4096 & 0.051 & Grid& 10 \\
& 356.73 & 1000 & 1024 & 0.821 & Grid& 10 \\
SiO (J=8$-$7) & 347.33 & 250 & 4096 & 0.053 & Grid& 20 \\
H$^{13}$CO$^{+}$ (J=4$-$3) & 347.00 & 250 & 4096 & 0.053 & Grid & 20 \\
\hline
\end{tabular}
\label{T:obserations_settings}
\end{table*} 

Mapping observations of $^{12}$CO, $^{13}$CO and C$^{18}$O (J=3$-$2) and single pointing grid observations of HCO$^{+}$, H$^{13}$CO$^{+}$ (J=4-3) and SiO (J=8-7) were obtained with HARP-B and the ACSIS autocorrelator \citep{Buckle2009} at the James Clerk Maxwell Telescope (JCMT), Mauna Kea, Hawaii on the 1$^{st}$ and 3$^{rd}$ of November 2009 as service proposal S09BU01. The spatial resolution of the JCMT at the observed frequencies is $\sim$14\arcsec{}, with a main beam efficiency $\eta_{MB}$~=~0.61 \citep{Buckle2009}. Though HARP-B consists of a 4$\times$4 pixel array of receivers, four of these pixels (H00, H01, H02 and H14) were not operational at the time of observing. For each of the CO lines, four sets of observations in position-switching raster mode were undertaken with a quarter array shift (29.1\arcsec{}) between each scan. A 90\deg{} scan direction change was also implemented between each pair of observations, after which the array was offset by an eighth (14.6\arcsec{}) in both the x and y plane of the array and the process repeated. The resulting basket-woven maps cover an area of 4\arcmin{}$\times$4\arcmin{} centred on G059.6331$-$00.1906 ($\alpha$~=~19$^{h}$43$^{m}$49.68$^{s}$, $\delta$~=~+23\deg{}28\arcmin{}38.3\arcsec{} and V$_{LSR}$~=~+25~\kms{}). The correlator settings and observing times are summarised in Table~\ref{T:obserations_settings}.

The data were initially reduced using the Starlink program \textsc{orac-dr}, however the broad wings of the outflow were within the baseline region determined by this routine. The data were therefore rebaselined and smoothed with a gaussian kernel to lower velocity resolution in order to reduce noise using python scripts. During the $^{13}$CO and C$^{18}$O observations, one of the HARP-B detectors (H04) had a particularly bad baseline, so after examination these data were re-reduced with this detector excluded. All observations were then converted to the main beam temperature scale (T$_{\rm mb}$).

\section{Determination of Outflow Properties}
\label{S:properties}

Previous studies of outflows \citep[e.g.][]{Bally1999,Beuther2002b,Stojimirovic2006} have tended to define global velocity windows, often by eye, to apply to all spectra containing outflow emission, or to a single spatially integrated spectrum. While this may work well for simple regions with only one outflow, using a single velocity window for all spectra across a region assumes that the cloud systematic velocity and line-width do not vary with position and that the outflow has a roughly similar velocity extent in all spectra. In the case of our observations this is not a safe assumption, as the systematic velocity and the full-width half-maximum (FWHM) both vary by $\sim$2\kms{} across the outflow region and the velocity extent varies strongly with position. We therefore set out below the method and criteria we have used to determine the velocity window and mass individually for each spaxel within which emission is attributed to the outflow.

\begin{figure}
\center
\includegraphics[width=0.38\textwidth]{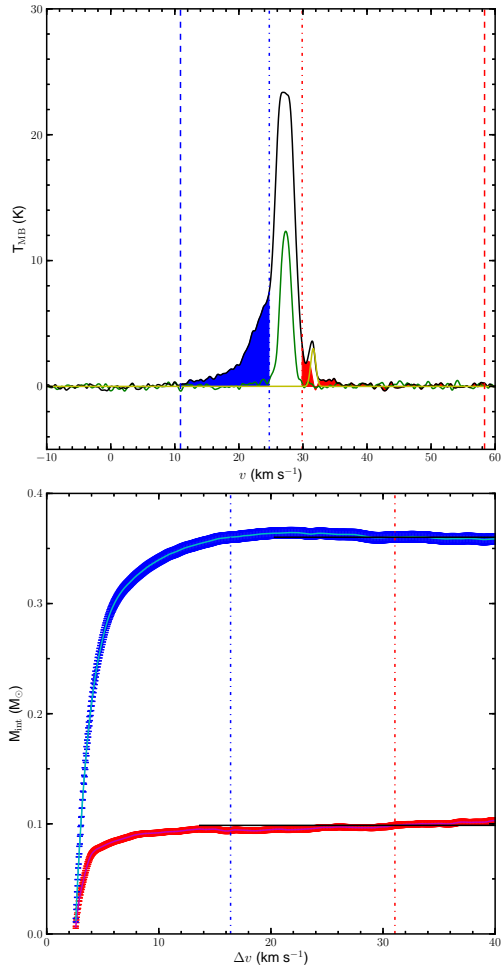}
\caption{Top: Example $^{12}$CO (black) and $^{13}$CO (green) spectra. The yellow line indicates a gaussian fit to a non-outflow emission feature, which was removed when calculating the outflow properties. The inner and maximum outflow velocities are shown by the dashed and dash-dotted lines respectively, with the $^{12}$CO spectrum between these velocities shaded the appropriate outflow wing colour. Bottom: Cumulative integral of the mass in the outflow wings versus velocity offset for the spectrum shown above. The error bars indicate the random error associated with the cumulative integrated mass. The total integrated mass, obtained as discussed in section~\ref{S:properties}, step~\ref{St:11} is indicated by the black line, while the velocity at which the cumulative integrated mass first becomes greater than or equal to the total mass, used as the maximum outflow velocity, is marked by the dot-dashed lines.}
\label{F:properties_spec}
\end{figure}

\begin{enumerate}

\item Some $^{12}$CO spectra show additional cloud emission at different systematic velocities to the main feature associated with the outflow (e.g. see Figure~\ref{F:properties_spec}). Gaussian fits were used to remove these, where possible, from the spectra before measurements of the outflow properties were performed. Strict acceptance criteria were used for the fit results so that the peak velocity and line-width did not vary too much from the expected values in order to ensure that these fits did not remove outflow emission. \label{St:1}\\

\item The maximum temperatures of both the $^{12}$CO and $^{13}$CO spectra were measured using gaussian fits to the central cloud emission. \label{St:2}\\

\item The excitation temperature (T$_{\rm ex}$) was calculated for the spectrum, assuming that the $^{12}$CO emission is optically thick and that the gas along a given line of sight can be characterised by a single excitation temperature. \label{St:3}\\

\item The $^{13}$CO optical depth ($\tau^{13}$) was calculated, assuming that the $^{13}$CO emission is optically thin. \label{St:4}\\

\item The systematic velocity of the cloud for the spectrum was taken to be the velocity of the peak of the gaussian fit to the $^{13}$CO emission line, as the $^{13}$CO spectra do not include a significant contribution from the outflow. \label{St:5}\\

\item The emission at velocities within 3$\sigma$ of the peak of the gaussian fit to the $^{13}$CO spectrum was associated with the cloud, rather than the outflow, in the $^{12}$CO spectrum. \label{St:6}\\

\item We estimate the $^{12}$CO to $^{13}$CO isotopic ratio as a function of velocity (R$_{12/13}$($v$)) for the spectrum, using a quadratic fit to the velocity window identified as cloud emission, in order to account for optical depth effects in the $^{12}$CO spectrum. The maximum value of the ratio was limited to 62 \citep{Langer1993}, the assumed intrinsic ratio when both transitions are optically thin. \label{St:7}\\

\item The $^{13}$CO column density per velocity channel ($N$($^{13}$CO)($v$)) was calculated from the $^{12}$CO spectrum, using R$_{12/13}$($v$) and $\tau^{13}$ to account for optical depth effects. \label{St:8}\\ 

\item The mass per velocity channel (m$_{H_{2}}$($v$)) was calculated from the $^{13}$CO column density per velocity channel, assuming that the ratio of the H$_{2}$ column density to the $^{12}$CO column density is 1.2$\times$10$^{4}$ \citep{Frerking1982}, that the ratio of the $^{12}$CO to $^{13}$CO column density is 62 \citep{Langer1993}, that the mean molecular weight of the gas is 2.8 \citep{Kauffmann2008}, and using data with a pixel size of 7.3\arcsec{}. \label{St:9} \\

\item The cumulative mass per velocity channel was calculated separately for the red and blue outflow lobes as a function of the velocity offset ($\Delta$$v$) from the systematic velocity of the cloud, starting at the inner outflow velocity defined in step~\ref{St:6} and stopping at a velocity offset of 57.5\kms{}, determined to be free of outflow emission in all spectra (see Figure~\ref{F:properties_spec} for an example). \label{St:10} \\

\item  We calculate the total integrated mass in the outflow lobe for the spectrum using a constant least-squares fit to the velocities larger than the point at which the $^{12}$CO spectrum first becomes negative, i.e. has roughly reached the background level. The maximum velocity of the outflow lobe for the spectrum is then defined as the velocity at which the integrated mass first becomes greater than or equal to the total integrated mass (see Figure~\ref{F:properties_spec} for an example). \label{St:11} \\

\item The momentum per velocity channel is calculated by multiplying the mass per velocity channel calculated in step~\ref{St:9} by the velocity offset of each channel from the systematic cloud velocity and the kinetic energy per velocity channel by multiplying by half the velocity offset squared. The total momentum and kinetic energy in each lobe for the spectrum are then calculated as the integrals of the momentum and kinetic energy per velocity channel between the inner and maximum outflow velocities, as defined in steps~\ref{St:6} and \ref{St:11}. \label{St:12} \\

\end{enumerate}

\begin{figure}
\center
\includegraphics[width=0.49\textwidth]{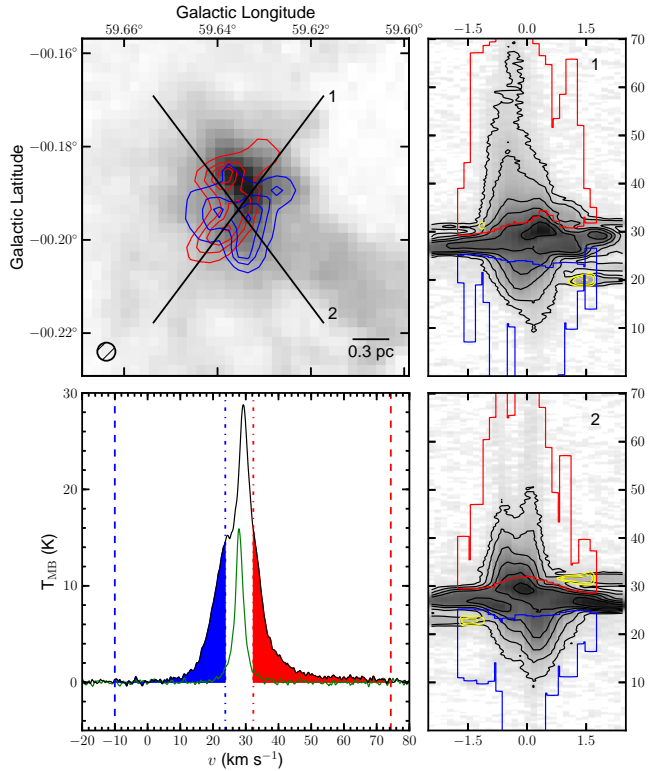}
\caption{Outflows identified from HARP-B data. Top left: Integrated $^{13}$CO emission shown in greyscale with red and blue contours showing outflow emission detected in $^{12}$CO for the same region as in Figure~\ref{F:intro_3colour}. The red and blue contour levels are 25, 50, 75, 100 and 125 K\kms{}. Right, top and bottom: Position-velocity (PV) cuts through the $^{12}$CO data, indicated by diagonal numbered lines in the top right plot, shown in a square-root scale to emphasise low-scale emission. The levels of the black contours to the $^{12}$CO data are 0.8, 2.15, 5.6, 10.5, 17 and 25\kms{}, while the yellow contours follow the same scale and indicate the regions of non-outflow emission removed using gaussian fits as discussed in the text. The red and blue contours indicate the red and blue outflow velocity windows. Bottom Left: Example $^{12}$CO (black) and $^{13}$CO (green) spectra for the pixel at the centre of the two PV cuts, with all lines and shaded regions having the same meaning as the top plot in Figure~\ref{F:properties_spec}.}
\label{F:properties_outflow1}
\end{figure}

While removing other cloud emission features in step~\ref{St:1} may have resulted in removal of a small amount of outflow emission, we consider such a conservative approach to be preferable to not removing this cloud emission. Examples of the regions where such emission has been identified for removal are indicated using yellow contours in the position-velocity plots in Figure~\ref{F:properties_outflow1}, while the outflow velocity windows are shown using red and blue contours respectively. The equations used for steps~\ref{St:3},\ref{St:4},\ref{St:8} and \ref{St:9} can be found in \citet{Wilson2009}.

While the random uncertainty in the mass due to thermal noise is relatively small (e.g. see the error bars in Figure~\ref{F:properties_spec}), these are undoubtably dominated by systematic errors. \citet{Buckle2009} estimate the systematic calibration errors for the JCMT to be $\sim$20$\%$ and the various abundance ratios used are not particularly well know. In addition, the values used here were primarily measured in average ISM/molecular cloud conditions, so may not be representative of the abundance in star formation regions where depletion, photodissociation and fractionation processes can vary depending on position, species and even isotope. Masking within 3$\sigma$ in step~\ref{St:6} is reasonably conservative, as shown in Figure~\ref{F:properties_outflow1}, and probably leads to some outflow emission at low velocity offsets being removed, perhaps resulting in as large as a factor of 3 underestimation of the outflow mass \citep[][]{Downes2007}. However, this approach is more accurate than either not correcting for ambient material, which can result in a significant overestimate, or attempting a more complex correction without sufficient information \citep[][]{Cabrit1990}. Despite our conservative masking of the cloud emission, it is also possible that some of the emission considered as associated with the outflow is actually due to local turbulent material. While it is difficult to estimate a numerical error related to these factors, by using abundance ratios which are commonly used by other authors in the literature, we expect that we are not significantly more susceptible than other studies. The properties presented below are probably accurate to a factor of a few.

\section{Results}
\label{S:results}

\subsection{Outflow properties}
\label{S:results_co}

\begin{figure*}
\center
\includegraphics[width=0.9\textwidth]{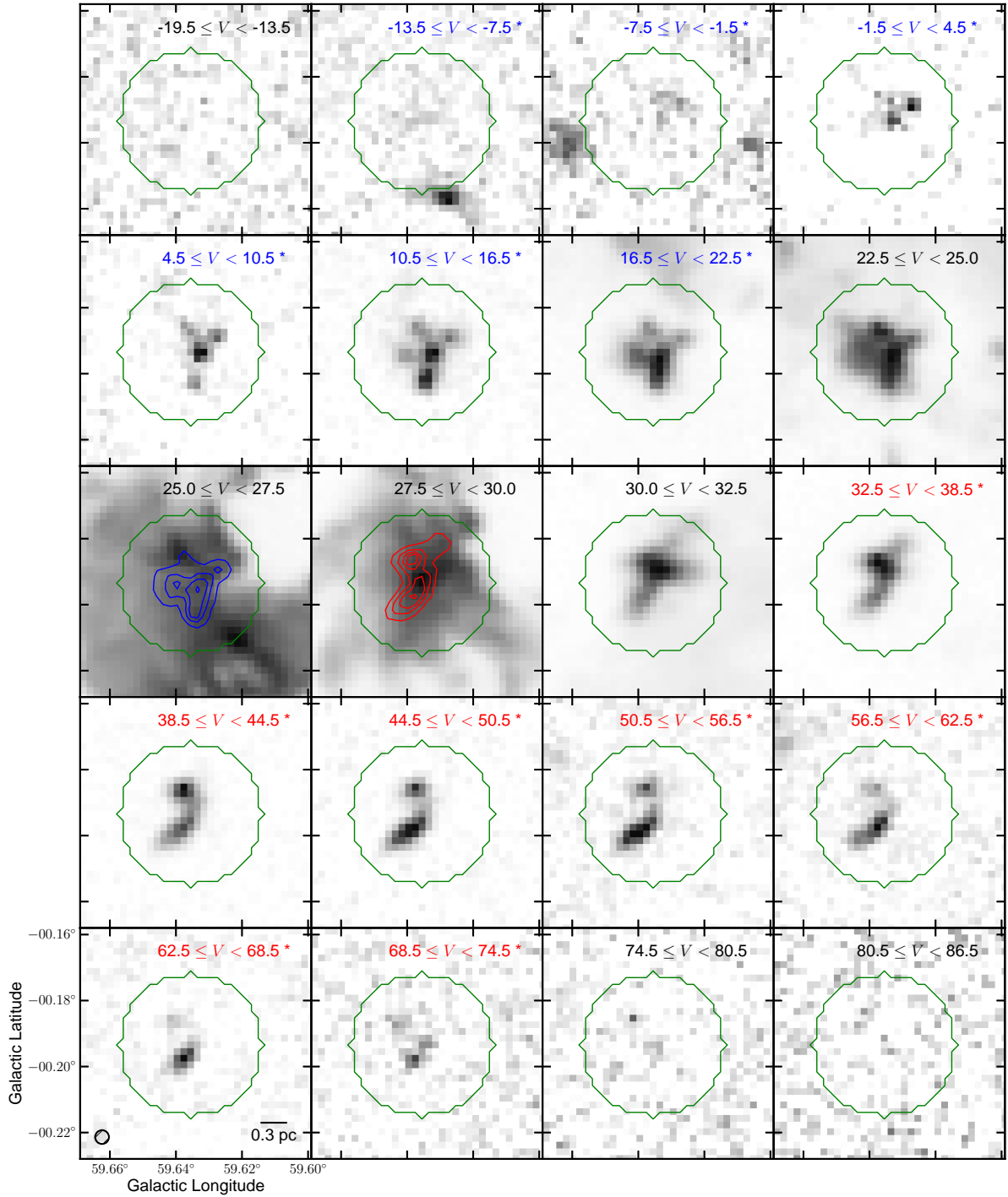}
\caption{HARP-B $^{12}$CO channel maps, integrated in 6~\kms{} slices except for the four slices at near the systematic velocity, which are 2.5~\kms{} in order to better reveal the exclusion of systematic emission from the outflow. The region to considered when calculating the total mass in the region is indicated by the green contour, while the velocity integration of each slice is given in the top right corner of each plot. Those slices where at least 1$\%$ of the pixels lie within the outflow velocity window are denoted with a star and the text is colour-coded to the outflow wing. Contours of the total integrated emission in each outflow lobe are overlaid on the central velocity slices using the same contour levels as in Figure~\ref{F:properties_spec}. The beam-size is shown in the bottom left of the plot.}
\label{F:results_channel}
\end{figure*}

We performed the steps outlined in section~\ref{S:properties} on all HARP $^{12}$CO spectra within a spatial radius of 10 pixels ($\sim$0.81pc) of the centre of the source, as this region encompasses the whole outflow. A series of channel maps with the channels identified as contributing to the outflow indicated are shown in Figure~\ref{F:results_channel}, with the region where our automated outflow identification was undertaken shown by the green contour.

\begin{table*}
\centering
\caption{Measured outflow properties}
\begin{tabular}{@{~}c|c|c|c|c|c|c|c|c@{~}}
\hline
\centering
Component & \multicolumn{2}{c|}{$M$} & \multicolumn{2}{c|}{$\Delta v_{\rm max}$} & \multicolumn{2}{c|}{$P$} & \multicolumn{2}{c}{$E$}\\
& \multicolumn{2}{c|}{(\msol{})} & \multicolumn{2}{c|}{(km~s$^{-1}$)} & \multicolumn{2}{c|}{(\msol{} km~s$^{-1}$)} & \multicolumn{2}{c}{(10$^{45}$ ergs)}\\
\hline
& r & b & r & b &r & b &r & b \\
\hline
1& 8.5& 30.3& 38.6 & 38.9 & 48.9& 196.3& 3.29& 15.93\\
2& 9.7& 26.0& 50.5 & 41.6 & 74.9& 142.1& 8.94& 9.50\\
3& 18.7& 5.8& 48.5 & 42.6 & 173.3& 44.8& 28.13& 4.51\\
4& 22.9& 6.9& 46.9 & 37.4 & 175.7& 52.9& 19.02& 4.90\\
\hline
total& 59.8& 69.1& & & 472.9& 436.1& 59.38& 34.84\\
\hline
\end{tabular}
\label{T:results_properties}
\end{table*} 

\begin{figure}
\center
\includegraphics[width=0.47\textwidth]{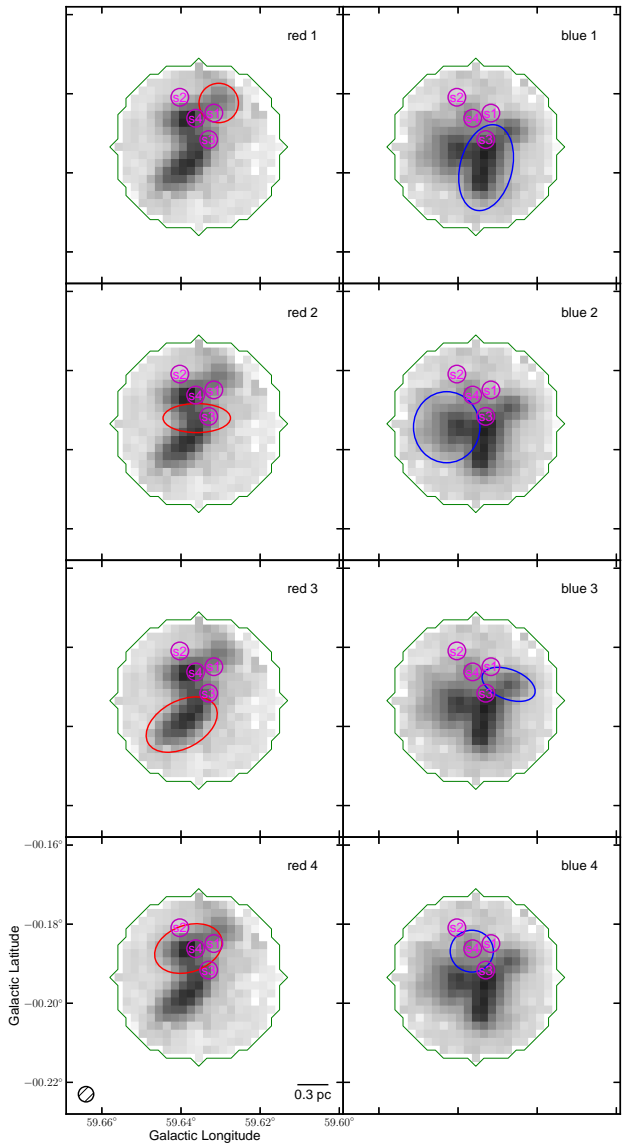}
\caption{Red and blue outflow components indicated with coloured ellipses overlaid on the integrated $^{12}$CO emission map corresponding to that line wing, as shown with contours in Figure~\ref{F:results_channel}. The numbers in the top right corner of each plot are the component numbers, also used in Table~\ref{T:results_properties}. The green contour is the same as in Figure~\ref{F:results_channel}. The four sources (s1-s4) associated with the clump containing the outflows are indicated in magenta. See sect.~\ref{S:results_seds} for more details.}
\label{F:results_outflow2}
\end{figure}

It is clear from Figures~\ref{F:properties_outflow1} and \ref{F:results_channel} that the single outflow originally identified in the FCRAO $^{12}$CO (J=1$-$0) data is at least partially resolved into multiple components at the resolution of the HARP-B observations. By examining both the channel and integrated maps, the blue wing emission breaks up into 4 or 5 components, while 3 or 4 separate components are visible in the red wing, though it is difficult to unambiguously identify the individual flows. In the following analysis, we tentatively identify four outflow components from the integrated emission maps in both the red and blue wings, as indicated in Figure~\ref{F:results_outflow2}, for which we have measured the mass, maximum relative velocity, momentum and kinetic energy, presented in Table~\ref{T:results_properties}. In regions of overlap, we assign spaxels in a mutually exclusive way. In order that our determination of individual flows does not overly affect our results, we also calculate the total properties of all the outflow emission. It is difficult to accurately estimate the angle of inclination between the line of sight and the outflow axis simply from our molecular observations, as the individual components are not well resolved or elongated, so we do not undertake an inclination correction for our velocities, momenta and energies. Thus the values of the quantities calculated below are lower limits to the true values and we are unable to calculate dynamical timescales.

\subsection{The Ambient Molecular Cloud}
\label{S:results_core}

The molecular clump associated with the outflow lies within a radius of $\sim$0.81~pc (10 pixel), shown by the green contour in Figure~\ref{F:results_channel}, and has a mass, measured from the $^{13}$CO observations, of 3655~\msol{}, with 458~\msol{} in dense gas as measured using the HARP-B C$^{18}$O (J=3$-$2) data and a ratio of the C$^{18}$O and $^{12}$CO column densities (i.e. $N$($^{12}$CO)~/~$N$(C$^{18}$O)) of 500 \citep{Frerking1982}. The same velocity window was used as for the outflow mass measurements, but without masking the line centre. Assuming that clump is spherical with a radius of 0.81~pc, we calculate its gravitational binding energy from the $^{13}$CO mass to be 1.4$\times$10$^{48}$~ergs. We also calculate the turbulent energy ($E_{turb}$) of the clump using:

\begin{equation}
E_{turb}~=~(3/16ln2)M_{cloud}\Delta V_{\textrm{FWHM}}^{2}
\end{equation}

from \citet{Arce2001} to be 1.7$\times$10$^{47}$~ergs, where $V_{\textrm{FWHM}}$ is the average $^{13}$CO FWHM of 2.4~\kms{} and $M_{cloud}$ is the $^{13}$CO clump mass. The combined energy of the outflow, even without inclination correction, is therefore approximately enough to drive the turbulence in the clump, and would provide enough energy to unbind the clump if i$\lesssim$15\deg{}.

\subsection{Associated YSOs}
\label{S:results_seds}

\begin{figure}
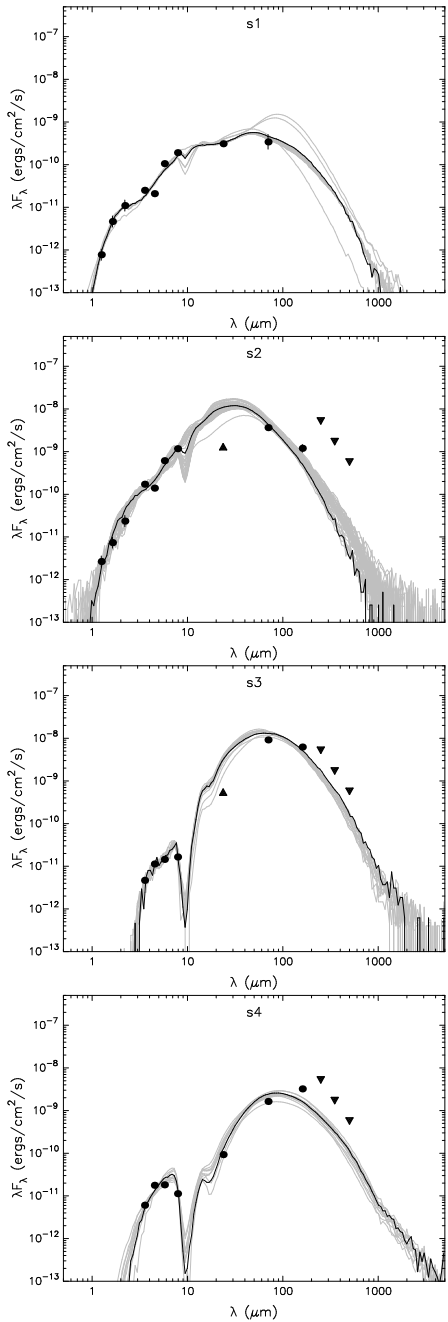

\center
\includegraphics[width=0.33\textwidth]{./s1.eps}
\includegraphics[width=0.33\textwidth]{./s2.eps}
\includegraphics[width=0.33\textwidth]{./s3.eps}
\includegraphics[width=0.33\textwidth]{./s4.eps}
\caption{SEDs of the four YSOs associated with the outflow material, located as indicated in Figure~\ref{F:results_outflow2}. Detections are indicated with filled black circled while upper and lower limits are indicated with upwards and downwards pointing filled black triangles respectively. The best fit model is shown as a black line while all other fits within a given $\Delta\chi^{2}$ of the best fit are shown by grey lines.}
\label{F:results_seds}
\end{figure}

In order to explore the (proto)-stellar sources which are potentially associated with the detected outflows, we first note that four sources can be seen in Figure~\ref{F:intro_3colour} which are probably associated with the molecular clump. Photometry data for these sources was gathered from the UKIDSS Galactic Plane Survey at $J$, $H$ and $K$ bands \citep[][]{Lucas2008}, the GLIMPSE survey in the four Spitzer IRAC bands \citep[][]{Benjamin2003} and preliminary PACS 70~\micron{} and 160~\micron{} from the Hi-GAL survey \citep[][]{Molinari2010b,Elia2010}. For the two sources not detected at all four IRAC bands, and for all sources from MIPSGAL 24~\micron{} images \citep[][]{Carey2009} aperture fitting photometry was performed to obtain flux data as in \citet[][]{Mottram2010a}. Sources 2 and 3 are mildly saturated in the 24~\micron{} images, so the fluxes for these sources are treated as lower limits. In the Hi-GAL SPIRE 250~\micron{}, 350~\micron{} and 500~\micron{} images, the sources are too close together for reliable photometry to be obtained for the individual components. We therefore performed one aperture fitting photometry measurement encompassing all the sources, and use the flux obtained as an upper limit for sources 2,3 and 4. Source 1 is not included in this process because it is not detected at 160~\micron{}. 

The spectral energy distributions (SEDs) of these sources were then fit using the model fitter of \citet{Robitaille2007} with similar input parameters to those used in \citet{Mottram2011a}. We use a distance of 2.3$\pm$0.5~kpc for all sources, where the distance error is set conservatively based on the difference between the photometric and kinematic distances rather than the error on the photometric distance from \citet{Massey1995}, which is much smaller. The SEDs and fits are show in Figure~\ref{F:results_seds}, while the results are presented in Table~\ref{T:results_seds}.

Despite the increase in resolution between the FCRAO and HARP-B data, it is still not simple to associate the sources directly with individual outflow components. However, sources 1, 3 and 4 all seem spatially related to the observed outflows, particularly sources 3 and 4. Based on values given in Table~1 of \citet{Mottram2011b}, the sources have luminosities consistent with spectral types in the mid to early B type, with the most massive (S3) being a B1$-$B0.5. The absence of detected radio continuum emission in Red MSX Source (RMS) Survey \citep{Urquhart2008} VLA 5~GHz observations of this region \citep{Urquhart2009} suggests that these sources are not powering \HII{} regions, so have yet to reach the main sequence.

\begin{table}
\centering
\caption{Properties derived from SED fitting}
\begin{tabular}{@{~}ccc@{~}}
\hline
\centering
Source & $F_{bol} (10^{-12} Wm^{-2})$ & $log_{10}(L/$\lsol{}$)$\\
\hline
s1& 3.2 $\pm$ 0.9 & 2.7 $\pm$ 0.2 \\
s2& 33.8 $\pm$ 14.4 & 3.8 $\pm$ 0.2 \\
s3& 50.8 $\pm$ 9.4 & 3.9 $\pm$ 0.2 \\
s4& 4.0 $\pm$ 0.7 & 2.8 $\pm$ 0.2 \\
\hline
total& 91.8 $\pm$ 17.2 & 4.2 $\pm$ 0.4 \\
\hline
\end{tabular}
\label{T:results_seds}
\end{table}

\subsection{SiO, HCO$^{+}$ \& H$^{13}$CO$^{+}$}
\label{S:results_chemistry}

\begin{figure}
\center
\includegraphics[width=0.48\textwidth]{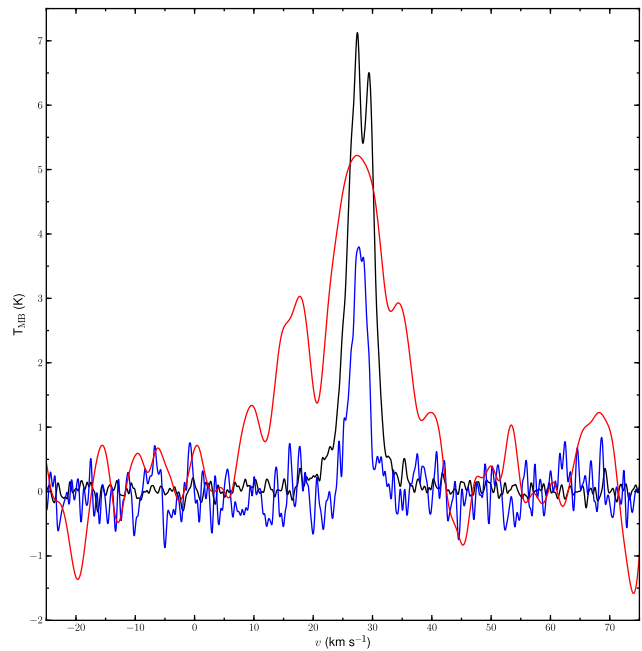}
\caption{HCO$^{+}$ (black), H$^{13}$CO$^{+}$ (blue) and SiO (red) spectra smoothed to 0.5~\kms{}, 0.5~\kms{} and 2.5~\kms{} respectively for the central position of the outflow. The H$^{13}$CO$^{+}$ spectra have been multiplied by 5 and the SiO by a factor of 25 for improved visibility}
\label{F:results_SiO_gridplot}
\end{figure}

The shock front and heating caused by molecular outflows interacting with surrounding cloud material strongly effects the chemistry of the surrounding region by releasing depleted molecules like SiO from grain mantles into the gas phase \citep[e.g.][]{Caselli1997,Schilke1997}, increasing it's relative abundance by factors of $\sim$10$^{2}$$-$10$^{6}$ with respect to the ambient material \citep[e.g.][]{Martin-Pintado1992,Garay1998}. Given that the depletion timescale for SiO onto dust grains is relatively short in normal molecular cloud conditions, of order 10$^{2}$$-$10$^{4}$~yrs \citep{Mikami1992,Martin-Pintado1992}, the presence of this molecule is therefore a good indicator that an outflow is currently active, rather than being a fossil flow. Our SiO observations are shown in Figure~\ref{F:results_SiO_gridplot}, where we detect broad emission (FWHM$\approx$20~\kms{}) towards the centre of the outflow, indicating that the outflows in this region are indeed active and not left over from previous activity. The amplitude observed is consistent with other observations \citep[e.g.][]{Klaassen2007}. The H$^{13}$CO$^{+}$ observations have a line-centre of 27.8~\kms{} and a FWHM of 3.3~\kms{}, consistent with the C$^{18}$O lines for the same position, so the HCO$^{+}$ and H$^{13}$CO$^{+}$ emission is most likely associated with the core/envelope rather than the outflow. The HCO$^{+}$/H$^{13}$CO$^{+}$ ratio for this location is in the range 7$-$28, much lower than the expected abundance ratio for $^{12}$C/$^{13}$C of 62 indicating that HCO$^{+}$ is optically thick and thus the double-peaked emission profile is almost certainly due to self-absorption.

\section{Discussion \& Conclusions}
\label{S:conc}

The total mass within the outflow components, as obtained from the HARP-B $^{12}$CO (J=3$-$2) observations is 129~\msol{}, while the core mass  derived from the BGPS 1.1mm dust continuum source is 327~$\pm$~112~\msol{}. The mass in dense gas, as measured from C$^{18}$O (J=3$-$2) data is 458~\msol{}, which agrees reasonably well with the core mass. 

Using the BGPS core mass and the relationship between outflow and core mass obtained by \citet[][their Fig~7.]{Beuther2002b} to the \citet[][]{Beuther2002a} sample results in an expected outflow mass of $\sim$10.3~\msol{}, an order of magnitude smaller than detected. The outflow-to-core mass ratio is 0.4, higher than any of the sources reported by \citet{Beuther2002a,Beuther2002b}, though the total luminosity of the associated YSOs is not particularly high compared to their sample. This suggests that these new outflows are either more efficient at entraining material or more energetic than those in the \citet{Beuther2002a} sample. If we assume an age for the outflows of 10$^{4}$~yrs (5$\times$10$^{5}$~yrs), the mechanical luminosity of the combined outflows is 79\lsol{} (2\lsol{}), resulting in a ratio of the mechanical to YSO luminosity of 0.5$\%$ (0.01$\%$). In order to compare with the results of \citet{Beuther2002b} we must first account for the increase which results from using the maximum outflow velocity and total mass rather than the channel velocity and mass when calculating the kinetic energy (approximately a factor of 25). Once this is taken into account, the highest mechanical to YSO luminosity ratio in their sample is 0.07$\%$ for IRAS 19410+2336 (G059.7833+00.0647 visible in the top left of Figure~\ref{F:intro_int12co}) and the median ratio is 0.002$\%$.

There are several factors which could result in such high outflow-to-mass and mechanical-to-source luminosity ratios compared with previous studies. Firstly, it is possible that the additional driving source(s) for the outflows are heavily embedded within dense dust and gas core, though if this were the case it is surprising that there is not even a suggestion of this in Figure~\ref{F:intro_3colour} as these data cover 3$-$250\micron{}. It could also be the case that despite our best efforts to be conservative, some cloud emission is still being included within the outflow velocity windows, but our mass measurements would have to drop by an order of magnitude to fall in line with other sources. The gas might have a higher than normal CO to H$_{2}$ ratio, leading to an overestimation of gas, though one has to ask why this particular region would be special in this regard. 

Finally, it could be that outflow activity from young massive stars is variable due to variable accretion, in a similar way to the senario suggested by \citet[][see also \citealt{Baraffe2010}]{Baraffe2009} for low-mass stars, where sources spend $\sim$1$\%$ of their time in a high-accretion (10$^{-4}$\msol\,yr$^{-1}$) phase and the rest accreting at a much lower rate (10$^{-6}$\msol\,yr$^{-1}$). If the outflow from such a source was observed during or soon after the high accretion phase, a much higher mechanical luminosity would be observed than after a long low-activity phase but the luminosity and core mass of the source would be similar. 

In summary, we have confirmed an outflow with multiple components, probably associated with a group of four YSOs, in the Vulpecula Rift which has an outflow mass of 129~\msol{} and a core mass of 327~$\pm$~112~\msol{}. The combined kinetic energy in the outflows (94$\times$10$^{45}$~ergs) is enough to drive turbulence in the local clump and potentially unbind the region, depending on inclination and the efficiency with which that energy is translated to the cloud. The outflows are not only energetic but also certainly active, due to the SiO detection, and thus presumably the related mid to early B type YSOs are currently still undergoing a phase of major accretion. Given the large reservoirs of cold dust detected in the sub-mm towards these sources, most likely associated with dense envelopes, this phase will probably continue for some time. The relatively red SEDs and the energetic nature of the outflows point to the relative youth of these sources, so it is likely these YSOs are destined to become more massive, rather than reach the main-sequence at their current mass.

Overall, some variation in outflow-to-core mass ratio is to be expected due to variations in cloud conditions, accretion rates, entrainment efficiency and the powering source. However, the outflow mass, momentum and energy must be in some way related to the mass and age of the central source, and thus the energy able to be injected, as found empirically by \citet{Beuther2002b}. To find a young region where the outflows appear to be an order of magnitude more efficient than expected warrants further detailed investigation, in order to explore how outflows entrain mass from and impart energy and momentum to the surrounding molecular material. We therefore intend to undertake further interferometric observations at higher spatial resolution to study this region.

\section*{Acknowledgments}

The authors thank the anonymous referee for comments and suggestions which improved the clarity of the paper and Jennifer Hatchell for many helpful discussions during the course of this work. We are grateful to Michele Pestalozzi and Sergio Molinari for provision of improved Herschel Hi-GAL photometry and images. We also thank Ian Coulson and the staff of the JCMT for their assistance before and during our observations, and Thomas Robitaille and Eli Bressert for help with plotting using python. This work was supported by STFC Grant ST/F003277/1 to the University of Exeter. The JCMT is operated by The Joint Astronomy Centre on behalf of the Science and Technology Facilities Council of the United Kingdom, the Netherlands Organisation for Scientific Research, and the National Research Council of Canada. The Five College Radio Astronomy Observatory was supported by NSF grant AST 0838222.

\bibliographystyle{./mn2e}
\bibliography{./MottramJC_vrs_outflow_2011_bib}

\label{lastpage}

\end{document}